\documentclass[twoside]{elsart}
\journal{Astroparticle Physics}
\usepackage{graphics}
\usepackage{epsfig}
\usepackage{graphicx}
\usepackage{rotate}
\usepackage{color}
\usepackage{natbib}
\usepackage{pifont}
\usepackage{amssymb}  % Mathesymbole
\begin{document}% HEADER
%Definitions
\def\intunits{\rm s^{-1}\,sr^{-1}\,cm^{-2}}
\def\diffunits{\rm GeV s^{-1}\,sr^{-1}\,cm^{-2}}
\def\fluxunits{\rm GeV^{-1} s^{-1}\,sr^{-1}\,cm^{-2}}
\def\diffcr{\frac{dN_{\rm CR}}{dE_{\rm CR}}}
\def\diffp{\frac{dN_{p}}{dE_{p}}}
\def\diffnu{\frac{dN_{\nu}}{dE_{\nu}}}
%DEFINITIONS:
% Energies:
\def\en{E_{\nu}}
\def\eg{E_{\gamma}}
\def\ep{E_{p}}
\def\ecr{E_{\rm CR}}
\def\ee{E_{e}}
% break energies:
\def\epb{\epsilon_{p}^{b}}
\def\enb{\epsilon_{\nu}^{b}}
\def\enbG{\epsilon_{\nu,GeV}^{b}}
\def\enbM{\epsilon_{\nu,MeV}^{b}}
\def\ens{\epsilon_{\nu}^{s}}
\def\ensG{\epsilon_{\nu,GeV}^{s}}
\def\egb{\epsilon_{\gamma}^{b}}
\def\egbM{\epsilon_{\gamma,MeV}^{b}}
\def\eauger{E_{\rm Auger}^{\min}}
\def\emincr{E_{\rm p}^{\min}}
\def\eminnu{E_{\rm \nu}^{\min}}
\def\lumi{L_{\gamma}^{52}}
%%%%%%%%%%%%%%%%%%%%%%%%%%%%%%%%%%%%%%%%%%%%%%%%
%  version March 30, 2008
%%%%%%%%%%%%%%%%%%%%%%%%%%%%%%%%%%%%%%%%%
\begin{frontmatter}
\title{Neutrinos from active black holes, sources of ultra high energy cosmic rays}
\author[gbg,dort,cor]{Julia K.~Becker}
\author[bonn,bonn2,alabama1,alabama2,karls]{and Peter L.~Biermann}
\address[gbg]{G\"oteborgs Universitet, Institutionen f\"or Fysik, SE-41296 
G\"oteborg, Sweden}
\address[dort]{Technische Universit\"at Dortmund, Inst.~f.~Physik, 
D-44221 Dortmund, Germany}
\address[bonn]{Max Planck Institut f.\ Radioastr., Auf dem
  H\"ugel 69, D-53121 Bonn, Germany}
\address[bonn2]{Dept.\ of Physics and Astronomy, University of Bonn,
  Germany}
\address[alabama1]{Dept.\ of Physics and Astronomy, University of
  Alabama, Tuscaloosa, AL, USA}
\address[alabama2]{Dept.\ of Physics and Astronomy, University of
  Alabama, Huntsville, AL, USA}
\address[karls]{Inst.~Nucl.~Phys.~FZ, Karlsruhe Inst.~of Techn.~(KIT), Karlsruhe, Germany}
\corauth[cor]{{\scriptsize Corresponding author. Contact: julia.becker@physics.gu.se, phone: 
+46-31-7723190}}
\date{\today}
\begin{abstract}
A correlation between the highest energy Cosmic Rays (above $\sim 
60$~EeV) and the distribution of active galactic nuclei (AGN)
gives rise to a prediction of neutrino production in the same
sources. In this paper, we present a detailed AGN model, predicting neutrino production near the
foot of the jet, where the photon fields from the disk and 
synchrotron radiation from the jet itself create high
optical depths for proton-photon interactions.
The protons escape from later shocks where the 
emission region is optically thin for proton-photon
interactions. Consequently, 
Cosmic Rays are predicted to come from FR-I galaxies,
independent of the orientation of the source. Neutrinos, on the other hand, are only
observable from sources directing their jet towards Earth, i.e.~flat
spectrum radio sources and in particular BL Lac type objects, due to the
strongly boosted neutrino emission.
\end{abstract}
\end{frontmatter}
\parindent=0cm
\clearpage
%%%%%%%%%%%%%%%%%%%%%%%%%%%%%%%%%%%%%%%%%%%%%%%%%
\section{The underlying AGN model\label{agn_model}}
%%%%%%%%%%%%%%%%%%%%%%%%%%%%%%%%%%%%%%%%%%%%%%%%%
The evidence for a correlation between the arrival 
directions of ultra high energy Cosmic Rays (UHECRs) with 
active galactic nuclei, as reported by the \cite{auger_science2007,auger_astrop2008}, supports many long standing expectations~\citep{ginzburg_syrovatskii1963}.  The active galactic nuclei with the most 
detailed available theory to actually accelerate protons to beyond 
$10^{20}$ eV are radio galaxies~\citep{biermann_strittmatter1987}. In all 
radio galaxies the feeding of outer radio emitting regions is done via a 
relativistic jet emanating from near a black hole.  Shock waves in such 
jets can accelerate particles just as shocks in the Solar wind do\footnote{For
a first discussion of the Solar wind, see~\citep{biermann1951}, for
shock acceleration in AGN see \citep[e.g.]{bednarz_ostrowski98}.}.
Shocks in the jet emanating from near the black hole start
around a few thousand gravitational radii, as it was shown with detailed
spectral fits of the entire electromagnetic
spectrum, including the spatial structure at the wavelengths where it is
known. The shocks end as strong shocks at
kpc or further out~\citep{markoff2001,markoff2005}. It was 
also shown that when particles get accelerated at the first shock,
proton-photon interactions limit their maximal energy
\citep{kruells1992,nellen,mannheimjet1995}.

The correlation with the distribution of active galactic nuclei
claimed by the Auger collaboration has been disputed by the HiRes
collaboration \citep{hires2008}. However, it is not clear at this point,
whether both data sets use the same lower energy cutoff with the same
sharpness. This is important, as the MHD simulations of cosmic
magnetic fields \citep{ryu2008,das2008} show that scattering of ultra high energy
particles rapidly increases with lower energy even near 60 EeV. Thus,
with even a slight mismatch between the two data sets the statistics
could be very skewed. Using a complete sample of radio galaxy sources
and their predicted properties as UHECR sources, these statistics will
be explored elsewhere \citep{caramete2008,curtiu2008,dutan2008}.
%=======================================================
\subsection{FR-I galaxies and UHECRs\label{unified}}
%=======================================================
Radio galaxies with extended radio jets were classified into two
categories by~\cite{fanaroff_riley1974}: A
population of high luminosity shows radio lobes at the outer edge of
the jet, at kpc scales from the core, {\it Fanaroff Riley II objects, short
FR-II}. The lower luminosity
population, on the other hand, has radio knots distributed along the
jet, the first knot being as close as $\sim 3000$~Schwarzschild radii
from the central core, {\it Fanaroff Riley I objects, short
FR-I}. Both radio lobes and knots show non-thermal radio spectra,
arising from electron acceleration at a shock front as first
theoretically described
by~\cite{fermi1949,fermi1954}. In analogy to processes in the creation
of Galactic Cosmic Rays, protons are believed to be accelerated
at those shock fronts in the same way as electrons. 
In particular, oblique shocks can be very efficient in particle
acceleration, and use electric fields in shock-drift acceleration due
to the Lorentz transformation of the magnetic fields in the proper
frame, see~\citep{hoffmann_teller1950,parker1958,jokipii1987,voelk_biermann1988,biermann1993,meli_biermann2006,athina_paper}.
%------------------------------------------------------------------
\subsubsection{Unified model of FR-I galaxies and BL Lac objects}
%------------------------------------------------------------------
In this paper, we consider FR-I galaxies as the sources of the UHECRs
potentially observed by Auger. It is discussed by
others~\citep{oana2003,cuoco_hannestad2008,halzen_omurchadha2008,koers_tinyakov2008,kachelriess2008}
that the nearby FR-I galaxy NGC5128, Centaurus A (Cen~A in the following), is
a good candidate to be responsible for a large fraction of the
correlated events above $60$~EeV. M~87 is another closeby candidate, see~\cite{biermann_strittmatter1987},
which cannot contribute to a possible Auger correlation, since it is barely in
Auger's field of view. The large number of more distant
FR-I galaxies provide good candidates for the total Cosmic Ray flux above
the ankle. Here, we discuss the morphology of FR-I type galaxies and
how these can accelerate particles to the highest energies.

\begin{figure}[h!]
\centering{
\epsfig{file=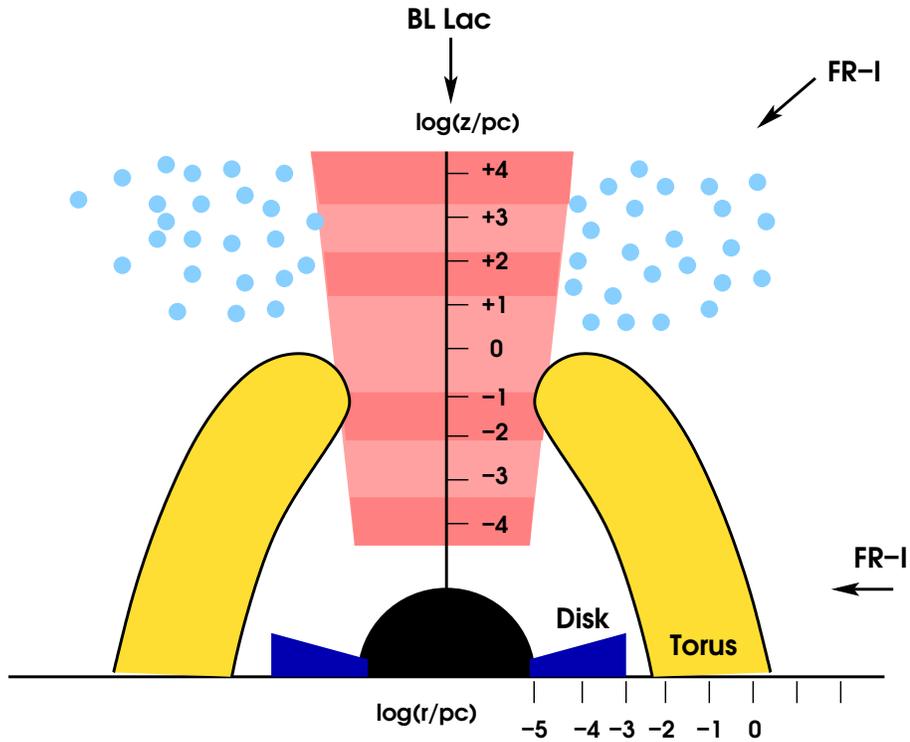,width=12cm}
\caption{Schematic figure of the class of FR-I galaxies with
  double-logarithmic scales. When the jet
 is pointed directly towards Earth, FR-I galaxies are classified as
  BL Lac objects. In this model, the torus will hide the accretion
  disk from view. In the case of a FR-I galaxy without torus,
  no radiatively efficient disk may be present.\label{agn_scheme_fri:fig}}
}
\end{figure}

Figure~\ref{agn_scheme_fri:fig} presents a schematic view of the model
of FR-I galaxies that we use in this paper. On the x-axis, the
rotational-symmetric part of the AGN is shown, while the y-axis
represents the axis of rotational symmetry along the AGN jet. Both
axes have logarithmic units. When
the AGN jet is pointed towards Earth, the FR-I type galaxy is viewed
as a BL Lac type object \citep{urry_padovani1994,urry_padovani1995},
showing flat radio spectra, with an unresolved jet
structure. When the AGN jet is
viewed at an angle, the jet structure with radio knots distributed
along the jet can be observed. In contrast to the more radio-luminous FR-II galaxies,
FR-I type objects typically lack the observation of optical
disks. While some of the FR-I type galaxies, such as M~87, clearly
lack luminous accretion disks\footnote{They are likely to have
  radiatively inefficient accretion disks, typically for low accretion
  rates.} and tori, many objects in this class may have accretion
disks hidden behind the torus, provided that the torus is closed
around the jet as indicated in Fig.~\ref{agn_scheme_fri:fig},
see~\citep{falcke_krishna_biermann1995}. In this case, the jet-disk symbiosis
model holds also for FR-I galaxies and the disk power scales with the
radio luminosity. 
%=======================================================
\subsubsection{Magnetic fields and shock structure in FR-I galaxies}
%=======================================================
The dependence of the magnetic field in these jets along the jet axis $z_j$ 
is near $B \, \sim \, z_{j}^{-1}$ at large distances. However, in the inner 
region, the radial dependence is not certain. Here, we investigate the
radial dependence of the magnetic field in order to determine the
Cosmic Rays' maximum energy along the jet and in particular, at the
innermost part of the jet.

Observations~\citep[e.g.]{bridle_perley1984}
suggest that the magnetic field runs as $z_{j}^{-2}$ at 
first, since the radio polarization observation show that the magnetic 
field is parallel to the jet. By analogy to the solutions of 
${\rm div} {\bf B} \, = \, 0$ in a magnetic wind~\citep{parker1958}, the 
magnetic field locally shows a parallel component of $B_{||} \, \sim \, z_{j}^{-2}$. Further out, 
the magnetic field observations suggest that, just as in a wind, the 
magnetic field becomes perpendicular to the flow direction, and so
$B \, \sim \, z_{j}^{-1}$. It is obvious that in a smooth wind, any 
component decreasing with $z_{j}^{-1}$ will ultimately win over a component 
running as $z_{j}^{-2}$.  As the equation of state is almost certainly 
relativistic~\citep[e.g.]{falcke1995a,falcke1995b}, the total pressure $P$
depends on the density as $P \, \sim \, {\rho}^{4/3}$, 
while in a conical simple jet the density $\rho 
\, \sim \, z_{j}^{-2}$, giving in near equipartition (the magnetic field 
pressure running with the total pressure) then 
$B \, \sim \,\sqrt{P(r)}\,\sim\, z_{j}^{-4/3}$.

However, repeated shock waves will reheat the material (see e.g.~\cite{sanders1983}, as well as Mach's original work in the 19th
century, \cite{mach_wentzel1884,mach_wentzel1885,mach1898}), and 
so we will assume that the Mach-number of the flow repeatedly comes back 
to the same value, while the jet flow velocity will stay approximately 
constant. Therefore going from crest to crest
\begin{equation}
P \, \sim \, \rho\,,
\end{equation} 
and so as a consequence 
\begin{equation}
B \, 
\sim \, z_{j}^{-1}\,.
\end{equation}  
This is consistent with the concept that the jet 
stays approximately conical. This argument is independent of the 
orientation of the magnetic field, and so the radio polarization 
observations are not in contradiction, but need then an interpretation 
as arising from highly oblique shocks, which emphasize magnetic field 
components parallel to the shock surface. Highly oblique shocks are only 
possible for high Mach-numbers, which again is consistent.

\begin{figure}[h!]
\centering{
\epsfig{file=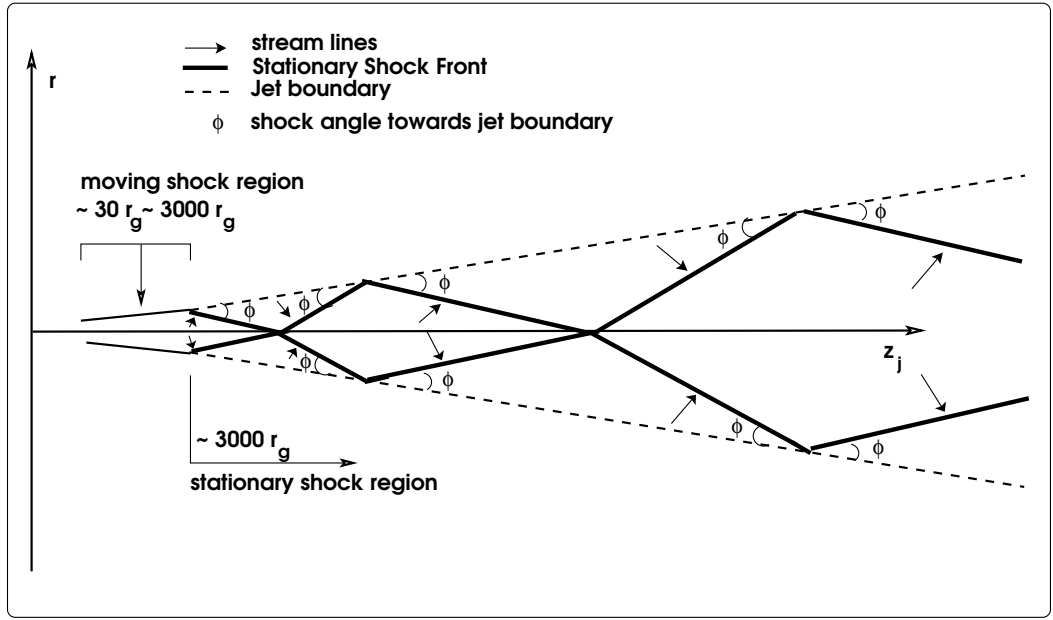,width=\linewidth}
\caption{Proposed conical shock structure in AGN jet.\label{agn_mach}}
}
\end{figure}
As one check let us consider a system of repeated conical shocks, and 
ignore for simplicity the inner Mach disks as shown schematically in
Fig.~\ref{agn_mach} (see also \cite{sanders1983}).  Then we can see, that given 
a specific Mach-number at the initial flow formation there will be highly 
oblique shock waves, which will repeatedly reflect on the conical 
boundaries, and so produce a self-similar pattern as long as the 
Mach-number keeps returning to near its initial value.  A self-similar 
repeated structure with an ever increasing inner scale will result.  
This is precisely what is seen in jet structure at vastly discrepant 
spatial resolutions, like in the radio galaxy NGC6251
(\cite{bridle_perley1984}, and earlier papers). Further examples for
such a shock structure are recent observations of BL
Lacertae~\citep{marscher2008} and the BL Lac type object S5~1803+784~\citep{britzen2008}.

For those oblique, stationary shocks, it is likely that a mixture of
sub- and superluminal shocks is present\footnote{The value of the angle between the
magnetic field and the shock front normal determines whether a
transformation into the Hoffmann-Teller frame (${\bf E}=0$) is
possible (subluminal) or not (superluminal).}. The proton spectra look
very different when comparing the two cases. While superluminal
spectra have maximum energies of around $10^{5}$~GeV, subluminal
shocks can reach energies up to the highest energies, i.e.~$E_{\max
}\sim 10^{21}$~eV, as shown by~\cite{athina_paper}.

Another consequence is that the magnetic field can be sufficiently high 
even far out to confine particles at the energies of UHECRs. As the magnetic field is anchored to the inner accretion 
rate, or its residual electric currents from an earlier accretion event, any 
magnetic field that decreases much faster than $z_{j}^{-1}$ on the way out, 
will have difficulty to confine particles during their acceleration to 
the highest energies observed~\citep{blandford1976,blandford_znajek1977,blandford_koenigl1979}.  Only a magnetic field which runs overall 
as $B \, \sim \, z_{j}^{-1}$ allows the magnetic field to be relatively high 
far out along the jet structure.

We conclude that the magnetic field then runs as $B \, \sim \,
z_{j}^{-1}$ approximately.
%--------------------------------------------------------------
\subsubsection{Cosmic Ray acceleration in FR-I galaxies}
%--------------------------------------------------------------
Normally when considering where UHECRs can be accelerated, the spatial limit, or Hillas-limit, is 
invoked~\citep{hillas1984}.
Using radio observations this suggests that radio hot spots of 
powerful Fanaroff-Riley II radio galaxies are very good bets: they are 
usually modelled as shocks~\citep{meisenheimer1989}, which can be 
shown to accelerate particles to near $10^{21}$ eV~\citep{biermann_strittmatter1987}.

When using the information on the magnetic fields inferred from 
radio jets, their radio knots, and their hot spots, one typically finds
$10^{-4}$ Gauss on kpc scales~\citep{miley1980,bridle_perley1984}.  On 
the other hand, the maximum magnetic field close to the black hole, so 
on scales such as a few Schwarzschild radii, is given by $10^{4} \, {\rm 
Gauss} \, (M_{BH}/10^{8} \, M_{\odot})^{-1/2}$~\citep{shakura_sunyaev1973,blandford1976,massi_kaufman2008}.  Therefore on radial scales over 
a factor of $10^{8}$ the magnetic field decreases by just this factor, 
quite indicative of a $r^{-1}$ behavior.  This was also used quite 
successfully by~\cite{blandford_koenigl1979}. Therefore, we conclude 
that such a radial dependence of the overall magnetic field is well 
justified.

When estimating what the magnetic field might be in a jet, we have 
used the current accretion rate to indicate an estimate, connecting it 
to the accretion disk~\citep{falcke1995a,falcke1995b}.
This then gives usually a rather weak field in all cases when the 
overall activity is low, such as believed in BL Lac objects.  And a weak 
field fails the Hillas test for UHECRs.

However, an alternative proposed by~\cite{blandford_znajek1977} is that the magnetic field is still quite high, 
from a prior accretion episode, and then the jet can be driven by a 
spin-down of the black hole. In this context the magnetic fields are 
higher, and even low power sources such as FR-I radio galaxies are 
possible sources of UHECRs~\citep{dutan_biermann2005,dutan_biermann2008}.

In addition, the Lovelace limit~\citep{lovelace1976} shows that 
the Poynting flux, a lower limit to the energy flux in a jet, is 
connected to the maximal energy of a particle confined in the jet by 
$L_{jet} \; = \, 10^{47} \, {\rm erg/s} \, (E_{\max}/10^{21} {\rm 
eV})^{2}$.  Therefore a jet such as Cen A, estimated to carry probably 
around $10^{43} \, {\rm erg/s}$ \citep{whysong_antonucci2003} cannot 
possibly accelerate particles - especially protons - to $10^{21}$ eV, at 
most it would seem, that $10^{19}$ eV is possible.  This limit can be 
exceeded by three arguments
 This limit can be exceeded through three
arguments:
(a) The particles might be heavier than hydrogen. In this case,
  however, photo-disintegration will reduce the neutrino flux, see
  e.g.~\cite{hooper2005,ave2005}. Hence, we only consider protons in our
  calculations and this argument does not apply here.
(b) Secondly and more importantly, \cite{gallant_achterberg1999} show that the Lorentz factor of
  the shock in the local upstream frame enters squared. 
(c) Further, the
  jets might be intermittent, as strongly demonstrated by the radio
  galaxy Hercules A \citep{gizani_leahy2003,nulsen2005}.

Therefore there is no a priori difficulty for FR-I radio galaxies to 
accelerate protons to near $10^{20}$ eV.
With a magnetic field decreasing as $z_{j}^{-1}$, the Hillas spatial limit will give the
same maximal proton energy at all radii. 
%============================================
\subsection{Optical depth}
%============================================
We discuss the interaction of ultra high energy cosmic rays with three
different targets along the AGN jet. We calculate the optical depth of
interactions with the photon fields of the disk and the knots'
synchrotron field. As a third possible target, we consider the
interaction of UHECRs where the jet meets the torus (see Fig.~\ref{agn_scheme_fri:fig}).
\begin{enumerate}
\item {\it Proton interactions with disk photons}\\
The optical depth for proton-photon interactions is given by the ratio of
the length $l\approx z_{j}\cdot \theta$ and mean free path of the
protons, $\lambda_{p\,\gamma_{\rm disk}}$, in the jet in the disk's photon field
$n_{\gamma_{\rm disk}}$:
\begin{equation}
\tau_{p\,\gamma_{\rm disk}}=\frac{l}{\lambda_{p\,\gamma_{\rm
      disk}}}=z_{j}\cdot \theta\cdot n_{\gamma_{\rm disk}}\cdot
  \sigma_{p\,\gamma}\,.
\end{equation}
Here, $\sigma_{p\gamma}=900\,\mu$barn is the total cross section for
the production of the Delta-resonance in proton-photon interactions~\citep{biermann_strittmatter1987}.
\parindent=0cm
\parskip=0.2cm

To check on a realistic value for $\tau_{p\,\gamma_{\rm disk}}$ we estimate the 
interaction probability as follows:

In an active galactic nucleus the accretion disk will produce a
radiation field near to the nucleus. The accretion disk 
luminosity at full efficiency is given by
\begin{equation}
L_{\rm disk}  \approx \epsilon_{Edd}\cdot  10^{44}  \, {\rm erg/s}\,,
\end{equation}
where $\epsilon_{Edd}\, < \, 1$ is the accretion rate relative to the 
maximum, the Eddington rate.

At some distance $z_j$ along the jet, starting at about 
$3000 \, r_g$, 
where $r_g \, = \, 1.5 \cdot 10^{12} \, M_{BH}/\left(10^{7} \, M_{\odot}\right) 
\, {\rm cm}$, the gravitational radius, we find stationary shock waves,
see~\citep{markoff2001,markoff2005} and
also~\cite{marscher2008}. Moving shock waves are expected to be
present between $\sim 10\,r_s-3000\,r_s$ as discussed in detail in~\cite{marscher2008}.  These shockwaves accelerate particles, and these particles, 
say protons, interact with the radiation field.  The photon density is 
then given by
\begin{equation}
n_{\gamma_{\rm disk}} =  \frac{L_{\rm disk}}{4 \pi \, z_{j}^{2} \, c \cdot h \nu}\,,
\end{equation}
where $h \nu$ is the typical photon energy, for which we adopt 20 eV, 
which is $h \nu \, = \, 3 \cdot 10^{-11} \, {\rm erg}$.  Using $3000 \, 
r_g$ as a reference radius, this expression can be rewritten as
\begin{equation}
n_{\gamma_{\rm disk}} \, = \, 4 \cdot 10^{11} \, \epsilon_{Edd}\cdot
\left(\frac{L_{\rm disk}}{10^{44}\,{\rm erg/s}}\right)\cdot \left({ \frac{z_{j}}{3000 \, r_g} }\right)^{-2} 
\, {\rm cm^{-3}}\,.
\end{equation}
The optical depth across the jet, with the length across the jet as
$\theta \, z_{j}$,  is the given by
\begin{equation}
\tau_{p\,\gamma_{\rm disk}} \, = \, 0.2 \cdot \, \epsilon_{Edd}\cdot
\left(\frac{\theta}{0.1}\right) \cdot  \left(\frac{z_{j}}{3000 \,
  r_g}\right)^{-1}\cdot \left(\frac{L_{\rm disk}}{10^{44}\,{\rm erg/s}}\right)\,,
\end{equation}
decreasing linearly outwards. 
Since FR-I galaxies are not as efficient radiators as FR-II galaxies
(see Section~\ref{unified}), the Eddington rate in the best case is
$\epsilon_{edd}\sim 0.1$. Those FR-I
sources with the torus close around the jet, the optical depth is
therefore $\sim 2\%$. For other FR-I type objects, like M~87, where
the accretion rate has decreased, the disk
can be come very faint. This means changing from a radiative disk,
\citep[e.g.]{shakura_sunyaev1973,novikov_thorne1973}, to a radiatively
inefficient disk, \citep[e.g.]{narayan_yi1995}, and the optical depth
will essentially be equal to zero. 
\item {\it Proton interactions with the synchrotron photon field in
  the jet}\\
 If protons interact with synchrotron photons in the same knot,
  the optical depth is given as
$\tau_{p\,\gamma_{\rm synch}}\approx z_{j}\cdot \theta\cdot
  n_{\gamma_{\rm synch}}^{rest}\cdot \sigma_{p\,\gamma}$.
The particle density of
synchrotron photons, $n_{\gamma_{\rm synch}}^{rest}$, is given in the
rest frame of the plasma and needs to be
transformed into the observer's frame, $n_{\gamma_{\rm synch}}^{obs}$. Even if the shock is
standing, the plasma is streaming with relativistic velocities along
the jet, and so is the synchrotron photon field. The photon density
observed at Earth is thus modified from the
field that the interacting protons 'see' in the plasma's rest frame,
see \cite[e.g.]{rybicki_lightman1979}:
\begin{equation}
n_{\gamma_{\rm synch}}^{rest}=\Gamma^{-1}\cdot
n_{\gamma_{\rm synch}}^{obs}\,.
\end{equation} 
Here, $\Gamma$ is the boost factor of the streaming plasma relative to Earth.
The photon density in the plasma's frame
can be determined by assuming that the luminosity
of a knot along the jet is a fraction $\epsilon_{\rm knot}\approx 0.1$ of the total
synchrotron luminosity, $L_{\rm synch}\approx 10^{40}$~erg/s:
\begin{equation}
n_{\gamma_{\rm synch}}^{obs}\approx \frac{\epsilon_{knot}\cdot L_{\rm
    synch}}{4\pi\cdot z_{j}^{2}\cdot c\cdot (h\nu)}\,.
\end{equation}
The frequency of
synchrotron photons is $\nu\sim
1$~GHz. 
Hence, the optical depth is 
\begin{eqnarray}
\tau_{p\,\gamma_{\rm synch}}\approx 0.9&\cdot& \left(\frac{10}{\Gamma}\right)
\left(\frac{\theta}{0.1}\right)\cdot\left(\frac{\epsilon_{\rm
    knot}}{0.1}\right)\nonumber\\
&\cdot& \left(\frac{L_{\rm synch}}{10^{40}\,{\rm
      erg/s}}\right)\cdot \left(\frac{z_{j}}{3000\,r_g}\right)^{-1}\cdot
  \left(\frac{\nu}{1\,{\rm GHz}}\right)^{-1}\,.
\end{eqnarray}
Thus, for a relativistically streaming plasma of $\Gamma\sim 10$,
optical depths around unity are expected. At the foot of the jet,
where it is still very collimated, it can therefore be expected that a
large fraction of the protons interacts before escaping, producing
neutrinos. Farther outside in the jet, the size of the knots increases
with the distance along the jet and the optical depth will
decrease. So, even in the case of synchrotron radiation, the main location of neutrino production is expected at the foot of the jet, while
protons will be able to escape the source at larger distances from the core.
\item {\it Proton interactions with the proton field}\\ 
For FR-I galaxies with closed tori, the outer edges of the jet will
pass through the torus, at a distance of $z_{j}\sim 1-10$~pc from the
central black hole. The column depth of the torus is\footnote{For a
  discussion of the torus' column depth in AGN,
  see~\cite{zier_biermann2002} and references therein.} $X \sim 4\cdot
10^{23}\,{\rm cm}^{-2}$. The
proton-proton optical depth is therefore
\begin{equation}
\tau_{p\,p_{\rm torus}}=X\cdot \sigma_{p\,p}\approx 2\cdot 10^{-3}\,,
\end{equation}
with $\sigma_{p\,p}\approx 50$~mb. This interaction
efficiency of 0.2\% is small compared to the proton-photon optical
depths discussed above. In addition, only the small fraction of all
protons in the outer parts of the jet will interact (see
Fig.~\ref{agn_scheme_fri:fig}). We conclude that proton-proton interactions can be
neglected here\footnote{This situation could change drastically for AGN with extremely high
matter densities, i.e.~GPS/CSS sources, where both proton-proton
interactions and pion-proton interactions can occur~\cite{odea}. Those sources are
not considered in this paper: due to their high proton-proton
interaction rate, they are not good candidates for the observed
charged Cosmic Ray spectrum. They could be very efficient neutrino
emitters.}.
\end{enumerate}
To sum up, proton-photon interactions are the dominant source of
high-energy neutrino production. Optical depths are around unity at
the foot of the jet and decrease with the distance from the
core. Therefore, we expect neutrino production in the first large
shock at around $z_{j}\sim 3000\,r_{g}$.
%============================================
\subsection{Consequence for neutrino and Cosmic Ray emission}
%============================================
As a consequence of the high optical photon-proton optical depth close
to the
foot of FR-I jets, there has
to be abundant neutrino production near that first shock, and that shock
is in the already high speed relativistic flow. On the other hand, as
we now suspect that there is lateral substructure in these relativistic
flows~\citep[e.g.]{krishna,lovelace1976}, things 
could be subtle as regards the specific beaming~\citep[e.g.]{lind_blandford1985}. This implies that the spatial
conditions suffice for the production of $10^{12}$~GeV protons at the 
base of the jet, but losses due to proton interactions with the disk's photon
field will take several powers of ten from that as we discuss in the following subsection. The 
ultra high energy particles
we do observe probably receive their last energy increase at the last
strong shock (see
\cite{biermann_strittmatter1987}), since the optical depth decreases
with the distance from the AGN core. 
Neutrino observations may have the 
spatial resolution to check on this.

\begin{figure}[h!]
\centering{
\epsfig{file=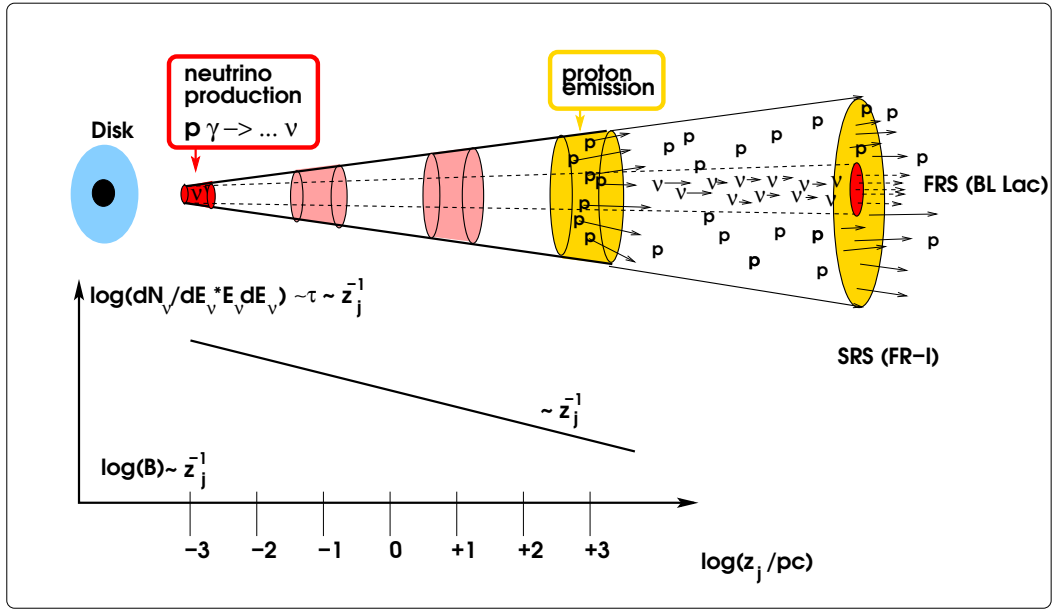,width=\linewidth}
\caption{Schematic figure of the AGN jet. The colored regions
  represent the emission regions. The detailed view of the shock
  structure discussed in Fig.~\ref{agn_mach} is not shown here, but it is
  assumed implicitly. While neutrino production happens
  in early, dense shocks, protons are more likely to come from the last
  shock, which is optically thin to proton-photon interactions. In this scenario,
  protons can be observed from FR-I galaxies, viewed
  from the side, and also from BL Lac objects, looking directly into
  the jet. Neutrino
  emission, on the other hand, is collimated and can only be observed 
from BL Lac objects. \label{agn_jet}}
}
\end{figure}
The schematic
view of this AGN model, showing the AGN jet with its shocks, is
displayed in Fig.~\ref{agn_jet}.
Since neutrino production happens close to the foot of the AGN jet in a
strongly accelerated reference system, the emission of neutrinos is
beamed. 
The neutrino background due to active galactic nuclei has been estimated 
many times, usually in conjunction to the process of accelerating ultra 
high energy particles, see e.g.~reviews
by~\cite{ghs1995,halzen_hooper2002,julias_review}. In particular, flat
spectrum radio quasars (FSRQs), interpreted as FR-II sources with
their jet pointing towards Earth, have been predicted to emit
neutrinos by~\cite{mannheim_stanev_biermann1992,atoyan_dermer2001,atoyan_dermer2003,bbr2005}.
Further, one extensive work is
by~\cite{bednarek_protheroe1999}, 
who consider the acceleration of protons very close to 
the disk, up to distances of $10^{16}$ cm, and using a reconnection 
model for the acceleration, a process well studied in activity regions 
on the Sun. In our model, on the other hand, we discuss distances of about $3 \cdot 10^{3}$ 
gravitational radii, so for typical black hole masses of $10^{8} \, 
M_{\odot}$, so distances somewhat larger, about $4 \cdot 10^{16}$ cm, 
and adopt the point of view that shock wave acceleration is the dominant 
process. On the other hand, just as~\cite{bednarek_protheroe1999}, we use 
 the disk radiation as the photon field for interaction.  Their figure 
3a demonstrates the limit of such an approach:  A typical disk 
temperature is of order $3 \cdot 10^{4}$
K~\citep{malkan1982,donea_biermann1996}, or sometimes even higher, as
also used in~\cite{bednarek_protheroe1999}.  
They confirm that the optical depth is of order 
unity. However, there is one critical difference: In shock 
acceleration the energetic protons are isotropic in the comoving frame, 
here moving with a Lorentz factor of order $10 - 30$, or perhaps even 
higher~\citep{gopal_krishna2007}.  So we do not have in our concept a proton moving 
rectilinearly straight out, but protons in a phase space distribution
moving along with the relativistic jet flow \citep{falcke1995a}. Hence, particle emission is beamed along the
jet axis.

Neutrinos are therefore only
observable from Earth if the jet points (almost) directly towards the
observer. Since protons are likely to be produced in the last strong
shock, AGN seen from the side also contribute to the flux of Cosmic
Rays.
%--------------------------------------------
\subsubsection{Flat Spectrum Radio Sources}
%--------------------------------------------
As a consequence of the beamed neutrino emission, neutrinos correlated 
to the proton emission of radio galaxies can only be observed from flat 
spectrum radio sources (FRS), i.e.~FR-I galaxies with their jet pointing
towards Earth. 

Most recently, the detection of a double-structure flare in optical
and X-ray wavelengths from the flat
spectrum radio source BL Lacertae was interpreted as the 
emission from particle acceleration in moving shock fronts very close
to the central black hole. Then, when the plasma reaches the turbulent
zone which marks the transition from moving to stationary shocks, a
second synchrotron flare is seen from particle acceleration in the
stationary shock, see~\cite{marscher2008} for details. During the
first X-ray flare, $>200$~GeV emission from BL Lacertae was detected by the
MAGIC telescope~\citep{albert_bllacertae}. \cite{marscher2008} explain
this high-energy component by Inverse Compton scattering of the accelerated electrons with
the synchrotron photons. An alternative explanation would be the
production of very high-energy photons via photohadronic
interactions. Here, charged and neutral pions are produced, the
charged pions decaying to produce neutrinos and the neutral pions
resulting in TeV photons. Thus, if the TeV signal is of hadronic
nature, i.e.~if it arises from $\pi^0-$decays, neutrinos are produced
simultaneously, as also discussed by \cite{mannheim_stanev_biermann1992}.
%--------------------------------------------
\subsubsection{Steep Spectrum Radio Sources}
%--------------------------------------------
Steep spectrum radio sources are AGN seen from the side, often
showing a detailed jet structure, with radio knots along the jets (FR-I
galaxies) or radio lobes at the outer end of the jet. We predict that
those sources are weak neutrino sources, considering the model
presented above: The beamed emission
from the jet is not directed towards Earth and very strongly focused
neutrino emission cannot be observed from Earth. 

As for one of the brightest radio
galaxies in the sky, Cen A, it 
will be possible to localize the origin
of neutrinos within the source: the extension of Cen A is ten
degrees on the sky, while the resolution of neutrino detection with IceCube
will be around $1^{\circ}$.  So, the 
flux of high energy neutrinos produced in each subsequent shock region  will
diminish with the distance from the black hole and the maximal energy will go up. However, as the jet of Cen A 
is not pointing near to the line of sight, we will see neutrinos only 
from secondary particles decay, after a primary charged particle has 
scattered in magnetic fields, near the boundary or outside the 
relativistic jet.  In this paper, we will calculate
the intensity of the neutrino signal connected to the possible 
correlation of UHECRs and the distribution of AGN as 
observed by Auger~\citep{auger_science2007,auger_astrop2008}.
%%%%%%%%%%%%%%%%%%%%%%%%%%%%%%%%%%%%%%%%%%%%%%%%%
\section{Estimate of the Cosmic Ray flux from nearby sources}
%%%%%%%%%%%%%%%%%%%%%%%%%%%%%%%%%%%%%%%%%%%%%%%%%
Recent Auger results~\citep{auger_science2007,auger_astrop2008} indicate that at least $20$ events above $57$~EeV are
correlated with the distribution of nearby AGN in the V{\'e}ron-Cetty
\& V{\'e}ron catalog, \cite{vcv2006}, abbreviated VCV catalog in the following.
The integrated exposure above $40$~EeV is given as Exposure~$=9\cdot 
10^{3}$~km$^2$~yr~sr and
can be assumed to be the same above $57$~EeV. Thus, the integral flux of
UHECRs is
\begin{equation}
N(>E)=\frac{\#(events)}{\rm Exposure}
\end{equation}
and in this case
\begin{equation}
N(>57{\rm EeV})=\frac{20}{9\cdot 10^{3}}\rm km^{-2} yr^{-1} sr^{-1}=7\cdot
10^{-21}\intunits\,.
\label{uhecr_int_auger}
\end{equation}
If the flux is correlated to a single or few point source(s) rather than to
many sources, the field of view (FoV) of Auger $\Omega_{\rm Auger}$ has 
to be taken into account in combination with the declination of the sources
$\omega_{\rm source}(\delta)$, as discussed in the approaches of~\cite{cuoco_hannestad2008,halzen_omurchadha2008,koers_tinyakov2008}. As opposed to those models, we calculate the diffuse 
contribution as a conservative estimate, based on a
detailed model concerning the physics of the AGN jet.
 
Based on the correlation claimed by Auger, we assume in the following
calculation that these $20$ events come from AGN in the supergalactic plane
(SGP). If the source population for the origin of the calculation is 
verified to be radio galaxies, it is likely that even more than $20$ events come from
AGN, since the VCV catalog is not complete.

The differential flux from the SGP can be assumed to follow a power-law,
\begin{equation}
\left.\diffcr\right|_{SGP}=A_{\rm
  SGP}\cdot\ecr^{-p} \,.
\label{uhecr_differential}
\end{equation}
Here, the cosmic ray energy $\ecr$ is given in the laboratory frame at
Earth. In all following calculations, energies $E$ are in the laboratory 
frame
at Earth, energies $E^{\rm source}=(1+z)\cdot E$ are given in the laboratory
frame at the source and $E'=E^{\rm source}/\Gamma=E\cdot (1+z)/\Gamma$
represent the energy in the shock rest frame, with $\Gamma$ as the bulk 
Lorenz
factor of the shock and $z$ as the cosmological redshift of the source. All
calculations are done in the way that energies have units of 
$[E]=[E^{\rm source}]=[E']=$~GeV.
The normalization factor $A_{\rm SGP}$ can be determined by using 
Equ.~(\ref{uhecr_int_auger}) and
comparing it to the integral form of Equ.~(\ref{uhecr_differential}):
\begin{eqnarray}
N(>\eauger)&=&\int_{\eauger}\,d\ecr\,\diffcr=A_{\rm
  SGP}\int_{\eauger}\ecr^{-p}\,d\ecr \nonumber\\
&\approx&A_{\rm SGP}\cdot(p-1)^{-1} (\eauger)^{-p+1}\label{uhecr_int}
\end{eqnarray}
Thus, the normalization factor $A_{\rm SGP}$ can be calculated to be
\begin{equation}
A_{\rm SGP}=N(>\eauger)\cdot(p-1)\cdot (\eauger)^{p-1}\label{uhecr_norm}
\end{equation}
with $\eauger$ in units of GeV and $[A_{\rm SGP}]=\fluxunits$.
The spectral index $p$ of UHECRs is observed to be $2.7$. However, 
stochastic shock
acceleration itself results in a spectrum with a spectral index of $2.3$ or
even flatter, see e.g.\ papers
by~\cite{bednarz_ostrowski98,kardashev1962,baring2004} or \cite{mbq_icrc2007,athina_paper}.
The steep
observed spectrum may be due to the fact that there is a distribution of
maximum energies or due to propagation effects. Therefore, we use index values 
of $p=2.3$
and $p=2.0$
for the sources of UHECRs in the SGP. For a
threshold energy of $57$~EeV and an integral flux of $7\cdot
10^{-21}\intunits$, the numerical value of the normalization constant is
\begin{equation}
A_{\rm SGP}=\left\{
\begin{array}{ll}
9\cdot 10^{-7}\,\fluxunits&\mbox{ for } p=2.3\\
4\cdot 10^{-10}\,\fluxunits&\mbox{ for } p=2.0\,.
\end{array}
\right.
\label{acr}
\end{equation}
\begin{figure}[h!]
\centering{
\epsfig{file=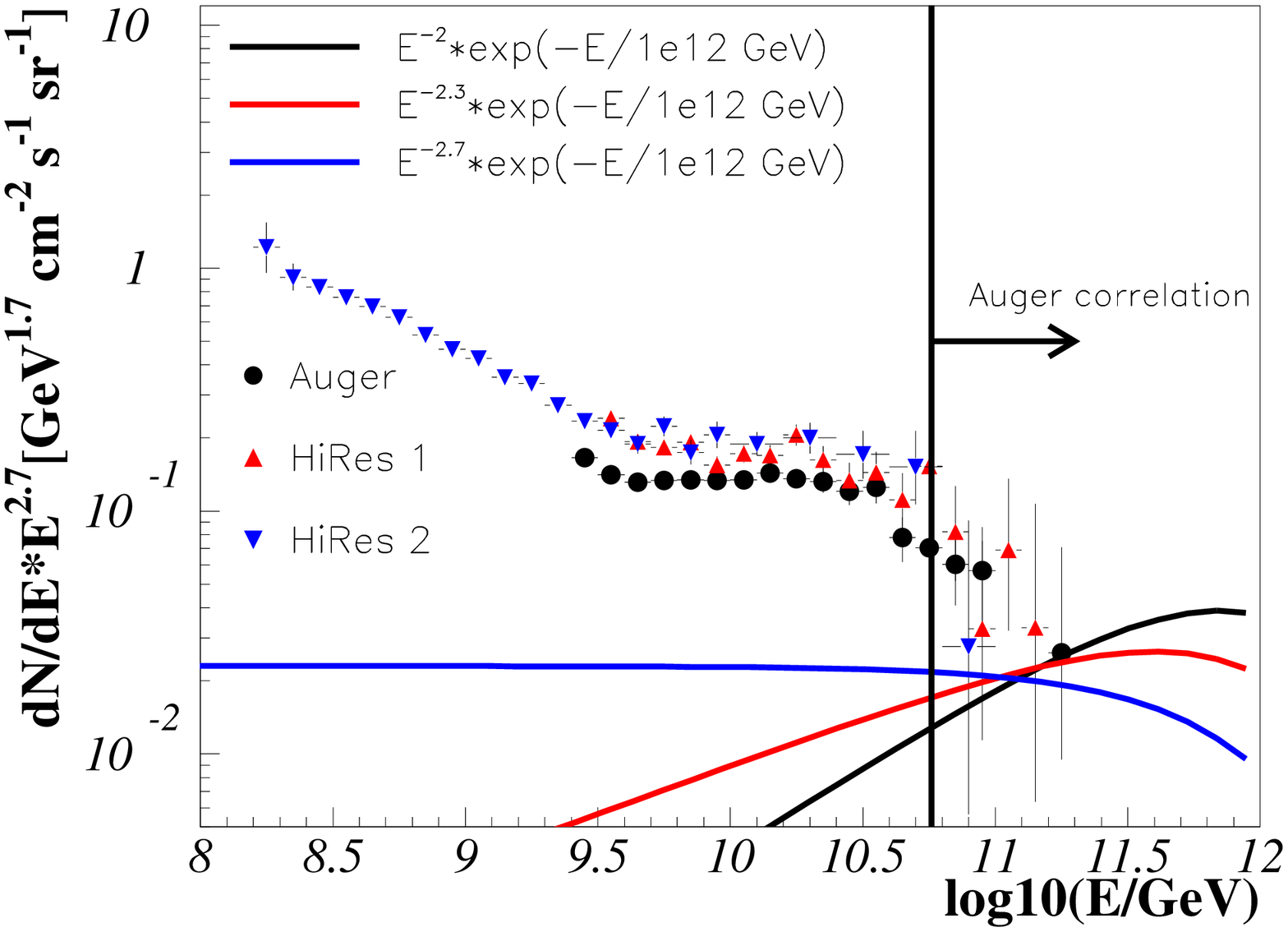,width=10.5cm}
\caption{Differential UHECR spectrum weighted with $E^{-2.7}$ - 
measurements from Auger (circles), \citep{auger_prl2008}, and HiRes-I
and HiRes-II (triangles), \citep{hires_prl2008}. The data from the two
different experiments are comparable when considering small
systematics in the energy measurements \citep{markus_roth_priv_comm}. The
  differential flux from the supergalactic plane calculated for $E^{-2}$
  (black line), $E^{-2.3}$ (red line) and $E^{-2.7}$ (blue line). An
  exponential cutoff was applied at $10^{12}$~GeV.\label{uhecrs_e27}}
}
\end{figure}
Figure~\ref{uhecrs_e27} shows the observed cosmic ray spectrum at UHEs,
weighted with ${\ep}^{2.7}$. Data are from the HiRes \citep{hires_prl2008} and
Auger \citep{auger_prl2008} experiments. The flux from the supergalactic plane as
calculated above is indicated as the blue ($p=2.7$), red ($p=2.3$) and black
($p=2.0$) lines. The three indices
represent different models for the description of Cosmic Ray
production and propagation following the theoretical results of
\cite{berezinsky2006} in the case of $p=2.7$, the work
of~\citep{bednarz_ostrowski98} for $p=2.3$, using parallel shocks, and
the results of \cite{mbq_icrc2007,athina_paper} for oblique shock
conditions ($p=2$). 
The energy flux  can be calculated from the differential spectrum as
\begin{equation}
j(E_{\rm th})=\int_{E_{\rm th}}\,E\,dN/dE\,dE\,.
\label{energy_flux}
\end{equation}
The total energy flux measured in cosmic rays
above $E_{\rm th}=10^{9.5}$~GeV is
approximately
\begin{equation}
j_{\rm tot-CR}\approx10^{-7}\,\diffunits\,.
\end{equation}
The corresponding energy flux from the supergalactic plane can be calculated
by using the same energy threshold and the differential spectrum determined
above:
\begin{eqnarray}
j_{\rm SGP-CR}&=&\left\{\begin{array}{ll}
\frac{A_{\rm SGP}}{p-2}\cdot E_{\rm th}^{-p+2}&\mbox{for }p\neq2\\
A_{\rm SGP}\cdot \ln(E_{\max}/E_{\rm th})\end{array}\right.\\
&\approx&\left\{
\begin{array}{ll}
4\cdot 10^{-9}\,\diffunits&\mbox{ for }p=2.3\\
9\cdot 10^{-10}\,\diffunits&\mbox{ for }p=2.0\end{array}\right.\,.
\label{energy_flux:numbers}
\end{eqnarray}
Here, a maximum energy of $E_{\max}=10^{10.5}$~GeV was used, taking into
account the GZK cutoff. This implies that about 1\% of the total cosmic ray
flux above the ankle is made up by sources in the supergalactic plane.
%%%%%%%%%%%%%%%%%%%%%%%%%%%%%%%%%%%%%%%%%%%%%%%%%
\section{Neutrino flux estimate from Auger measurements}
%%%%%%%%%%%%%%%%%%%%%%%%%%%%%%%%%%%%%%%%%%%%%%%%%
With protons being accelerated in AGN jets, neutrinos can be produced in
photohadronic interactions (see e.g.~\cite{julias_review} for a review),
\begin{equation}
p\,\gamma\rightarrow
\Delta^{+} \rightarrow \left\{\begin{array}{lll} \pi^{0}\,p \rightarrow 
\gamma\,\gamma\,(p)&&2/3 \mbox{ of the cases}\\
\pi^{+}\,n\rightarrow \mu^{+}\,\nu_{\mu}\,(n) \rightarrow 
\overline{\nu}_{\mu}\,\nu_{e}\,\nu_{\mu}\, (e^{+}\,n)&&1/3 \mbox{ of the 
cases}
\end{array}\right.\,.
\end{equation}
Here, the branching ratio for charged pion production is $1/3$. About 
$1/2$ of the
pion's energy goes into the sum of muon and anti-muon neutrinos
(Notation: $\nu:=\nu_{\mu}+\overline{\nu}_{\mu}$). In addition, neutrinos
oscillate on their way to Earth from a ratio of
\begin{equation}
(\nu_{e},\,\nu_{\mu},\,\nu_{\tau})|_{\rm source}=(1:2:0)\mbox{ to }
(\nu_{e},\,\nu_{\mu},\,\nu_{\tau})|_{\rm
  Earth}=(1:1:1)\,,
\end{equation}
see e.g.~\cite{stanev_book} and references therein.
Given an optical depth $\tau_{p\,\gamma}$
for the production of the Delta-resonance, the total neutrino energy
flux $j_{\nu}$, radiated in the solid angle $\Omega_{\nu}$,
$(\Omega_{\nu}\, j_{\nu}')=\int \en'\,dN_{\nu}/d\en'\,d\en'$, and the
total proton energy flux, focused within the solid angle $\Omega_{\rm
  CR}$,
$(\Omega_{\rm CR}\, j_{p}')=\int \ep\,dN_{p}/d\ep\,d\ep$, both given in the shock rest 
frame, are therefore connected as
\begin{equation}
(\Omega_{\nu}\cdot j_{\nu}')=\tau_{p\,\gamma}\cdot
\frac{1}{3}\cdot\frac{1}{2}\cdot \frac{1}{2}\cdot (\Omega_{\rm CR}\, j_{p}')=
\frac{\tau_{p\,\gamma}}{12}\cdot (\Omega_{\rm CR}\,j'_{p})\,.
\label{eflux_nu_p}
\end{equation}
The optical depth will be set to $\tau_{p\,\gamma}=1$ in the
following calculations (see Section~\ref{agn_model} for a detailed discussion). This correlation can be used to estimate the 
neutrino
flux to be expected from the supergalactic plane, given the  proton flux at
Earth as in Equations~(\ref{uhecr_differential}) and (\ref{uhecr_norm}). 
First,
we need to connect the energy flux in the shock rest frame and the one
measured at Earth.
%=======================================================
\subsection{The energy flux}
%=======================================================
The diffuse energy flux of a certain particle species is given in terms 
of the
energy spectrum at Earth as described in Equ.~(\ref{energy_flux}). In 
terms of
the single source spectrum, it can be written as
\begin{equation}
j=\int_{E^{\rm
    source}}\int_{z_{\min}}^{z_{\max}}\int_{L_{\min}}^{L_{\max}}dE^{\rm
  source}\,dz\,dL\,E^{\rm source}\,\frac{dN}{dE^{\rm source}}\cdot
\frac{1}{4\pi\,d_{L}^{2}}\cdot \frac{d^2n}{dV\,dL}\cdot \frac{dV}{dz}\,.
\end{equation}
Here, we are taking into
account the following facts:
\begin{itemize}
\item[(a)] Each source contributes with $E^{\rm source}\,dN/dE^{\rm
  source}\,dE^{\rm source}$.
\item[(b)] The flux of each source decreases with $4\,\pi\,d_{L}^{2}$, where $d_{L}$ is the 
source's luminosity distance.
\item[(c)] The source number per comoving volume $dV/dz$ and per luminosity
  interval, $dn/dV/dL$ - for protons, we use the radio luminosity function 
of FR-I
  galaxies, observed neutrinos are only produced in FRS.
\item[(d)] Sources up to a redshift $z_{\max}$, with an absolute upper limit
  of $0.03$ as the outskirts of the supergalactic plane contribute to
  the energy flux, both for neutrino and for proton sources. The
  minimum redshift is given by the distance of the closest plausible source, Cen~A, at
  $z_{\min}^{\rm CR}=0.0008$ in the case of FR-I galaxies\footnote{Cen
  A is located at a distance of $3.5$~Mpc which corresponds to a corrected
  redshift of $0.00083$ for a cosmology of
  $\Omega_m=0.3,\,\Omega_{\Lambda}=0.7$ and $h_0=0.7$, using
  approximate values from most recent WMAP-5year results~\citep{wmap_5yr}. The apparent, measured
  redshift is much higher than the actual cosmological one, due to the
  gravitational influence of the Great Attractor
  on nearby sources.} (CR sources), and Perseus A at
  $z_{\min}^{\nu}=0.018$ in the case of FRS (Neutrino sources).
\item[(e)] The luminosity integration limits for AGN have been chosen to 
match
  the observed distribution of FR-I galaxies, $L_{\min}=10^{40}$~erg/s and
  $L_{\max}=10^{44}$~erg/s. This applies for FRS as well, as these
  are generally believed to be a sub-class of FR-I galaxies.
\end{itemize}
As we would like to compare the diffuse energy flux at Earth to the energy
flux in the shock rest frame, we use $E^{\rm source}=\Gamma\cdot E'$ and 
receive
\begin{eqnarray}
j&=&\frac{\Gamma}{4\pi}\cdot\int_{E'}dE'\,E'\,\frac{dN}{dE'}\cdot\int_{z_{\min}}^{z_{\max}}\int_{L_{\min}}^{L_{\max}}dz\,dL\frac{1}{d_{L}^{2}}\cdot
\frac{d^2n}{dV\,dL}\cdot \frac{dV}{dz}\\
&=&\frac{\Gamma}{4\pi}\cdot j'\cdot n\,,
\label{j_source_prop}
\end{eqnarray}
with
\begin{equation}
n=\int_{z_{\min}}^{z_{\max}}\int_{L_{\min}}^{L_{\max}}dz\,dL\frac{1}{4\pi\,d_{L}^{2}}\cdot
\frac{d^2n}{dV\,dL}\cdot \frac{dV}{dz}\,.
\end{equation}
Inserting Equ.~(\ref{j_source_prop}) into 
Equ.~(\ref{eflux_nu_p}) for both neutrinos
and protons, we have
\begin{equation}
j_{\nu}=\frac{1}{12}\frac{\Gamma_{\nu}}{\Gamma_{\rm CR}}\cdot 
\frac{\Omega_{\rm
    CR}}{\Omega_{\nu}}\cdot\frac{n_{\nu}}{n_{\rm CR}}\cdot j_{\rm CR}\,.
\label{jnu}
\end{equation}
%================================================
\subsection{Discussion of parameters\label{parameters:sec}}
\begin{figure}[h!]
\centering{
\epsfig{file=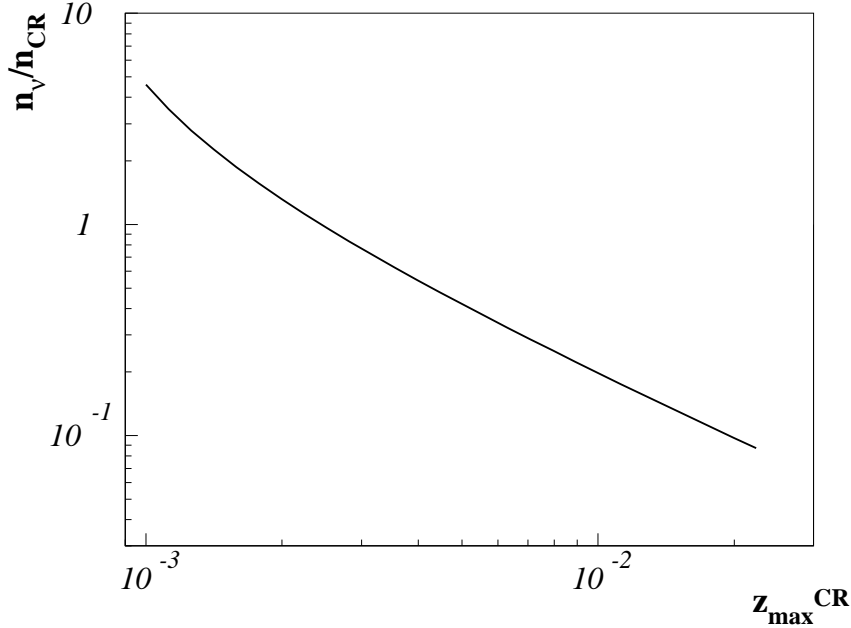,width=12cm}
\caption{Redshift factor $n_{\nu}/n_{CR}$ for with variable maximum redshift
for $n_{CR}$. If only very nearby sources are responsible for the 
correlation
of UHECRs with the distribution of AGN, $z_{\max}^{CR}$ can be as small as
$z_{\max}^{\rm CR}=0.001$. If the correlation comes from the entire distribution of
sources, the maximum redshift is given by the extension of the supergalactic
plane, $z_{\max}^{\rm CR}=0.03$. The minimum redshift for integration is 
taken to be
$z_{\min}^{\rm CR}=0.0008$ for the cosmic ray factor, using Cen~A as the
closest contributing source, and the neutrino factor has
$z_{\min}^{\nu}=0.018$ as Perseus~A as the closest contributing
source. The maximum redshift for neutrinos is taken to be $z_{\max}^{\nu}=0.03$. \label{nzmax}}
}
\end{figure}

%================================================
\begin{itemize}
\item {\em Cosmic ray energy flux at Earth}\\[0.2cm]
The proton energy flux can be written as
\begin{eqnarray}
j_{\rm SGP-CR}&=&A_{\rm SGP} \int d\ecr \ecr^{-p+1}\\
&=&A_{\rm Auger}\cdot \left\{\begin{array}{ll}
(p-2)^{-1}\cdot
({\ecr}^{\min})^{-p+2}&\mbox{ for } p\neq 2\\
\ln\left(\frac{E_{\rm CR}^{\max}}{E_{\rm CR}^{\min}}\right)&\mbox{ for } 
p=2\end{array}\right.
\\
&=& \frac{p-1}{p-2}\cdot N(>E_{\rm Auger}^{\min})\cdot (E_{\rm
  Auger}^{\min})^{p-1}\cdot (E_{\rm CR}^{\min})^{-p+2}\mbox{ for }p\neq 2\,.
\label{eflux_ap}
\end{eqnarray}
The proton energy is given in units of GeV and
$[A_{p}]=\fluxunits$.
For protons, the lower energy threshold is
given as the proton mass boosted by $\Gamma_{\rm CR}$, $E_{\rm
  CR}^{\min}=m_p\approx\Gamma_{\rm CR}\cdot 1$~GeV.\\[0.2cm]
\item {\em Neutrino energy flux at Earth}\\[0.2cm]
It is  assumed that the neutrino spectrum traces the proton
spectrum,
\begin{equation}
\diffnu=A_{\nu}\cdot \en^{-\alpha_{\nu}}
\end{equation}
with $\alpha_{\nu}\approx p$, $\en$ in units of GeV and
$[A_{\nu}]=\fluxunits$.
The neutrino energy flux can then be expressed as
\begin{eqnarray}
j_{\nu}&=&A_{\nu} \int_{E_{\nu}^{\min}} d\en
\en^{-\alpha_{\nu}+1}\\
&=&A_{\nu}\left\{
\begin{array}{ll}
(\alpha_{\nu}-2)^{-1}\cdot
(E_{\nu}^{\min})^{-\alpha_{\nu}+2}\approx 10\cdot
\Gamma_{\nu}^{-\alpha_{\nu}+2}&\mbox{ for }\alpha_{\nu}\neq2\\
\ln\left(\frac{E_{\max}^{\nu}}{E_{\min}^{\nu}}\right)&\mbox{ for }\alpha_{\nu}=2\,.
\end{array}\right.\label{eflux_anu}
\end{eqnarray}
The lower energy
threshold of one fourth of the pion mass, boosted by $\Gamma_{\nu}$,
$E_{\nu}^{\min}=\Gamma_{\nu}\cdot m_{\pi}/4=\Gamma_{\nu}\cdot 
0.035$~GeV, dominates and the upper
energy threshold can be neglected in the case of $\alpha_{\nu}\neq 2$. For
$\alpha_{\nu}=2$, the logarithm for neutrino energy equals approximately the
one for proton energies and the factors cancel in the end.

\parindent=0cm
\parskip=0.2cm
Concerning the threshold energies and spectral behavior of the
flux, our model differs from other approaches by~\cite{cuoco_hannestad2008,halzen_omurchadha2008,koers_tinyakov2008} in the sense
that we assume proton interactions with the photon field from the disk
and from the synchrotron radiation in the jet. In
the models of \cite{cuoco_hannestad2008,halzen_omurchadha2008,koers_tinyakov2008}, X-ray photons are assumed to interact with the
protons in the sources, which results in a broken power-law behavior
when calculating the neutrino flux. The reason is that the optical
depth for proton-photon interaction changes with energy, and this
change happens at a neutrino energy of around $\sim 10^{6}$~GeV for
X-ray photons. In the case of optical or radio photons, the break is at very
low energies, so that it is not relevant for our calculations. Thus,
the neutrino flux calculated here follows a single power-law just as
the proton spectrum does. 
\\[0.2cm]
\item {\em Lorentz factor}\\[0.2cm]
As neutrinos come from early shocks near the black hole, the beaming 
factor is
typically stronger than for protons from late, outer shocks. We will
assume in the following that the boost factor for neutrino production is a
factor of $\sim 3$ higher than for protons,
\begin{equation}
\frac{\Gamma_{\nu}}{\Gamma_{\rm CR}}\approx 3\,.
\end{equation}
\item {\em Solid angle}\\[0.2cm]
We assume in the following that $\Omega\propto \theta^2\propto 1/\Gamma^2$ as the typical
opening angle for relativistic sources. In particular, $\Omega_{\nu}\approx
1/\Gamma_{\nu}^{2}$ and $\Omega_{\rm CR}\propto 1/\Gamma_{\rm
  CR}^{2}$.\\[0.2cm]
\item {\em Redshift dependence}\\[0.2cm]
The redshift dependence of the neutrino energy flux is given as
\begin{equation}
n_{\nu}=\int_{z_{0.018}}^{z_{\max}^{\nu}}\int_{L_{\min}}^{L_{\max}}dz\,dL\frac{1}{4\pi\,d_{L}^{2}}\cdot
\left.\frac{d^2n}{dV\,dL}\right|_{\rm FRS}\cdot \frac{dV}{dz}\,.
\end{equation}
In a first approach, we will calculate the total diffuse neutrino flux from
the supergalactic plane and use $z_{\max}^{\nu}=0.03$ as the upper 
integration
limit, see also \cite{das2008}.
The radio luminosity function for flat spectrum radio sources,
$\left.d^2n/dV/dL\right|_{\rm FRS}$, determined
by~\cite{dunlop}. The authors do not distinguish between the
high- and low-luminosity sources, so both FR-II type objects
(``FSRQs'') and FR-I (``BL Lacs'') are included. Since the
low-luminosity part of the sources makes up most of the population, we
neglect the contribution of FSRQs here and interpret this function
of flat spectrum radio sources (FRS) as an approximation of the FR-I component of blazars. 

As discussed before, 
neutrino
emission is beamed and originates from FRS only. For protons, all FR-I
galaxies contribute:
\begin{equation}
n_{\rm CR}=\int_{0.0008}^{z_{\max}^{\rm 
CR}}\int_{L_{\min}}^{L_{\max}}dz\,dL\frac{1}{4\pi\,d_{L}^{2}}\cdot
\left.\frac{d^2n}{dV\,dL}\right|_{\rm FR-I}\cdot \frac{dV}{dz}\,.
\end{equation}
The radio luminosity function of FR-I galaxies is given in~\cite{willott}.
As the Auger correlation at most reveals a general correlation between the
distribution of AGN and cosmic rays, the exact number of sources is not
known. Note that the extension of sources goes with the distance
squared rather than with the total volume due to the disk-like
structure of the SGP, see~\cite{cavaliere_menci1997}. This is compensated by using the volume integration
for both the neutrino and the Cosmic Ray sources: The relevant ratio
$n_{\nu}/n_{\rm CR}$  gives the correct result.

Figure~\ref{nzmax} shows how the ratio $n_{\nu}/n_{\rm CR}$ 
depends on
the maximum redshift of contributing sources to the observed 
correlation. The
more sources contribute, the larger $n_{\rm CR}$ and the smaller the
ratio. The most conservative assumption is to assume that all sources in the
supergalactic plane
contribute to the correlation, $z_{\max}^{\rm CR}=0.03$, see also \cite{das2008}. The most optimistic
assumption is that only the nearest sources contribute, $z_{\max}^{\rm
  CR}=0.001$. Thus, the values lie between
\begin{equation}
0.1<\frac{n_{\nu}}{n_{\rm CR}}<5\,.
\end{equation}
\end{itemize}
~\vspace{0.5cm}

%=======================================================
\subsection{The diffuse neutrino flux from the supergalactic plane}
%=======================================================
Inserting the results from Section~\ref{parameters:sec} into 
Equ.~(\ref{jnu}) yields a numerical value for the
neutrino normalization constant,
\begin{eqnarray}
A_{\nu}&=&\frac{n_{\nu}}{n_{\rm
    CR}}\cdot\left(\frac{\Gamma_{\nu}}{\Gamma_{\rm
    CR}}\right)^{5-p}\cdot\frac{\tau_{p\,\gamma}}{12}\nonumber\\
&\cdot&
  N(>E_{\rm CR}^{\min})\cdot(p-1)\cdot(\frac{m_{\pi}}{4})^{p-2}\cdot
  (\eauger)^{p-1}\,,
\end{eqnarray}
where $\tau_{p\,\gamma}=1$ was assumed, as discussed in more detail in Section~\ref{agn_model}.

\begin{figure}[h!]
\centering{
\epsfig{file=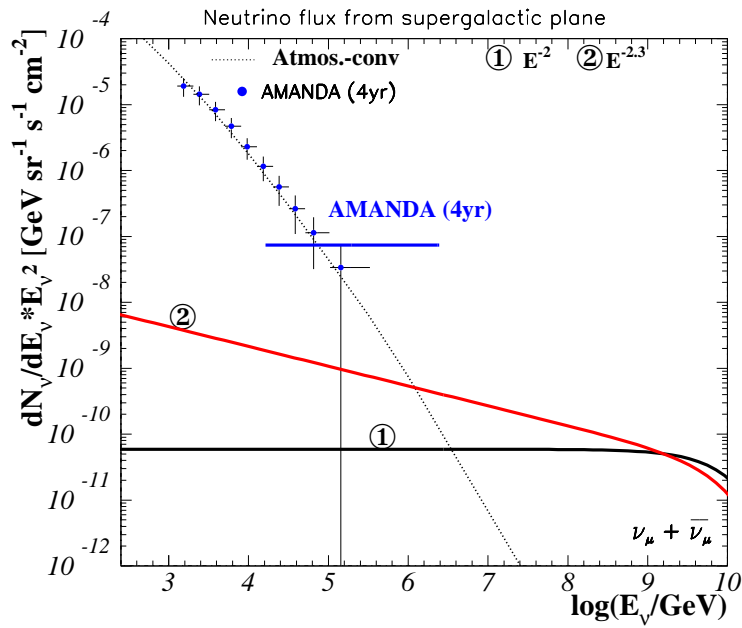,width=10.5cm}
\caption{Diffuse neutrino flux from the supergalactic plane, based on Auger
  data. The black line, 
labeled \ding{172}, is based on an $E^{-2}$ spectrum,
  the red line, labeled \ding{173}, is calculated using $E^{-2.3}$. AMANDA
  data are taken from~\cite{kirsten_icrc07} and the AMANDA limit is 
given in~\cite{jess_diffuse}. \label{sgp_neutrinos:fig}}
}
\end{figure}

\begin{figure}[h!]
\centering{
\epsfig{file=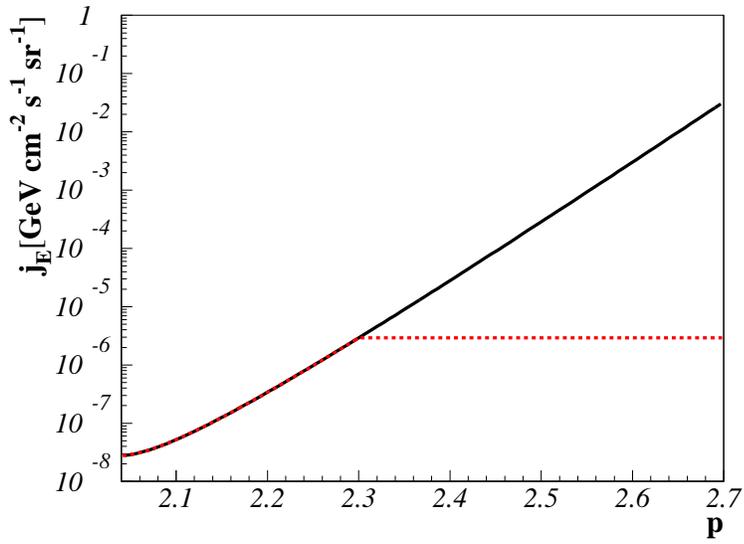,width=10.5cm}
\caption{Proton energy density, based on an $E^{-p}$ spectrum (black
  line). Assuming that a spectrum steeper than $E^{-2.3}$ at the highest
  energies actually approaches $E^{-2.3}$ at low energies, the energy 
density
  saturates at $p=2.3$ (red, dashed line). Still, there is a difference in
  the energy density of two orders of magnitude between assuming an $E^{-2}$
  spectrum and an $E^{-2.3}$ spectrum. \label{energy_density:fig}}
}
\end{figure}

The neutrino flux for an $E^{-2}-$ and an $E^{-2.3}-$ shaped spectrum is 
shown
in Fig.~\ref{sgp_neutrinos:fig}, assuming that the Auger correlation is 
caused
by all sources up to $z_{\max}^{\rm CR}=0.03$. The neutrino flux is 
higher at lower energies for the steeper spectrum
${\en}^{-2.3}$. The reason is that the normalization is done at Auger's
threshold energy. If the spectrum is steep $\sim {\en}^{-2.3\rightarrow-2.7}$, the 
energy content at lower
energies is higher. If the spectrum is very flat (e.g.\ $\en^{-2}$), the
contribution at energies lower than $E_{\rm CR}^{\min}$ is also lower (see
Fig.~\ref{energy_density:fig}). If the spectrum flattens towards smaller 
energies, this
reduces the normalization as well.

The numerical value of the
neutrino normalization constant for an $\en^{-2}-$spectrum is
\begin{equation}
A_{\nu}=\left\{\begin{array}{ll}
1.4\cdot 10^{-10}\,\diffunits&\mbox{ for }z_{\max}^{\rm CR}=0.03\\
5.0\cdot 10^{-9}\,\diffunits&\,\mbox{ for } z_{\max}^{\rm CR}=0.002\,.
\end{array}
\right.
\end{equation}
The expected neutrino flux is about a factor of $35$ higher if the 
closest AGN
produce the correlation of cosmic rays and AGN.
\clearpage
%=======================================================
\subsection{The total, extragalactic diffuse neutrino flux}
\begin{figure}[h!]
\centering{
\epsfig{file=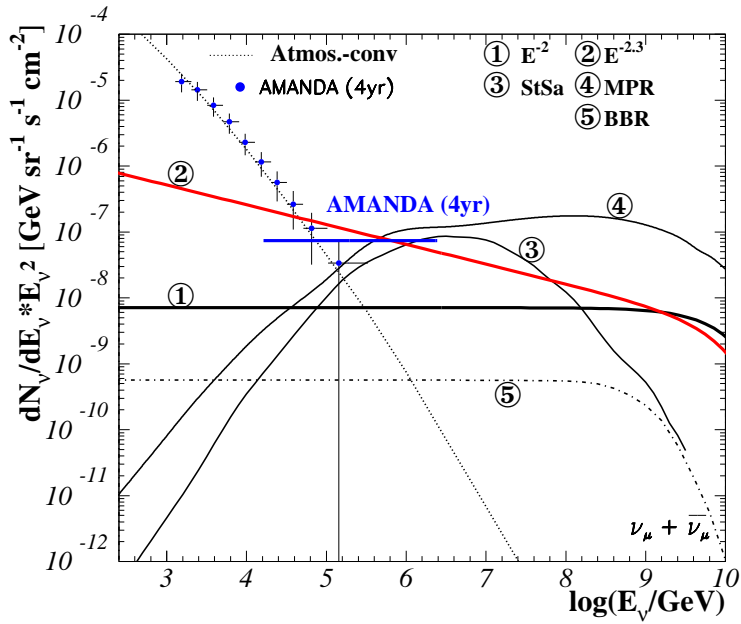,width=10.5cm}
\caption{Diffuse neutrino flux from flat spectrum radio sources up to $z<7$.
 The black line, labeled \ding{172}, is based on an $E^{-2}$ spectrum,
  the red line, labeled \ding{173}, is calculated using
  $E^{-2.3}$. The lines labeled \ding{174} and \ding{175} are shown
  for comparison, from~\cite{stecker_mod} and \cite{mpr},
  respectively. The two models use the high-energy component from AGN, $E_{\gamma}>$~MeV,
  to calculate the correlated neutrino flux. AMANDA
  data are taken from~\cite{kirsten_icrc07} and the AMANDA limit is given
  in~\cite{jess_diffuse}. The dashed line, labeled \ding{176}, shows the prediction of the
  neutrino flux from FSRQs as calculated in~\cite{bbr2005}. \label{tot_flux:fig}}
}
\end{figure}

%=======================================================
The total, extragalactic neutrino flux can be calculated by integrating 
up the
neutrino flux redshift dependence up to $z_{\max}^{\nu}=7$, where the first
active galaxies are believed to contribute, see~\cite{bouwens06,iye06}. The result is shown in
Fig.~\ref{tot_flux:fig}. Here, we use the conservative assumption that 
sources
up to $z_{\max}^{\rm CR}=0.03$ contribute to the correlation between cosmic
rays and the distribution of AGN. For an $E^{-2}-$shaped spectrum
(black, horizontal, solid line), the flux
is about a factor of $10$ higher than the prediction made in~\cite{bbr2005}, where the jet-disk
symbiosis model was used to estimate the neutrino flux from
FSRQs. Since we normalize the AGN spectrum at the highest energies, a
steeper neutrino spectrum (e.g.~$E^{-2.3}$, red, solid line) leads to
a higher flux at low energies. Although the figure seems to indicate
that an $E^{-2.3}-$ flux is already excluded by AMANDA data, it should
be noted that the limit is derived for an $E^{-2}-$ shaped
spectrum. To be able to exclude the $E^{-2.3}-$ spectrum, the limit
needs to be calculated for the same spectral behavior, see~\citep{julia_madison06,jess_diffuse}.
%=======================================================
\subsection{Uncertainties in the determination of the spectrum}
%=======================================================
Neutrino flux calculations typically bear three significant sources of
uncertainties, the first one being in the total normalization of the
spectrum, the second one lying in the uncertainty of the spectral
behavior. The third one is the maximum neutrino energy of the source class. Both quantities always rely on the internal properties of the
source class, which are typically poorly determined. This is true in
general and does not only apply to these calculations. Here, we
explain how these calculations are effected by the uncertainties.
\begin{enumerate}
\item {\it The normalization}\\
The uncertainty in the normalization of neutrino spectra has three main
components: the {\it measured spectrum} used to normalize the neutrino
flux, the {\it optical depth} in
the sources and the {\it opening angle} of neutrino and proton emission. \\
The correlation of the highest energy Cosmic Rays with the
VCV catalog is based on $20$ events. These statistics need to be
enhanced in order to achieve a more precise prediction of the actual
Cosmic Ray flux from the source class. In addition, the composition of
the spectrum is important for the neutrino flux, since protons produce
more neutrinos than heavy nuclei do. \\
Secondly, the optical depth of the
sources depends on the size of the acceleration region, on the
luminosity, on the boost factor and on the photon density in the
acceleration region. Those properties are known for a few single
objects, but they can vary with the source and also with time in a
given source.\\
The ratio of opening angles of proton and neutrino emission is
conservatively taken to be $\sim 3$. However, the real ratio clearly
depends on the properties of the single AGN, on the location of the
first and last shocks.
\item {\it The spectral index}\\
Depending on the orientation of the shock towards the magnetic field
and the boost factor,
the spectral behavior can vary. For parallel shocks, a spectral
behavior of $E^{-2.3}$ is expected as discussed
by~\cite{bednarz_ostrowski98}. If using large angle scattering instead
of pitch angle scattering for parallel
shocks, particle
spectra of up to $E^{-1.5}$ can be obtained \cite{stecker2007}. For oblique, subluminal shocks, the
spectra behave as $E^{-2.0\rightarrow -1.5}$, depending on the
boost factor, as discussed by~\cite{athina_paper}. 
\item {\it Maximum energy}\\
As discussed before, the maximum energy of the protons, and hence of
the neutrinos, depends on intrinsic quantities like the magnetic
field, boost factor and the disk luminosity. 
\end{enumerate}
So, to conclude, the importance of large volume neutrino detectors is
enhanced by the fact that protons or high-energy photons cannot give
unambiguous evidence for the spectral index, nor for the strength of
the neutrino flux. In order to identify the nature of the shocks in
not only AGN, but any Galactic or extragalactic accelerator, neutrinos
are essential. In this paper, we predict the {\emph region} of neutrino
production in AGN jets. We predict that flat spectrum radio sources
should be dominant. This can easily be tested by future experiments like IceCube and
Km3NeT. 
%%%%%%%%%%%%%%%%%%%%%%%%%%%%%%%%%%%%%%%%%%%%%%%%%%%
\section{Summary and implications}
%%%%%%%%%%%%%%%%%%%%%%%%%%%%%%%%%%%%%%%%%%%%%%%%%%%
In this paper, we present a model for Cosmic Ray and neutrino emission
from active galactic nuclei. A first evidence for the correlation of
the observed UHECRs above $60$~EeV and the distribution of nearby AGN gives
rise to the prediction that UHECRs may come from FR-I
galaxies. Although FR-II
galaxies would be a good candidate due to their high intrinsic
luminosity, they are not abundant enough and too far away to lead to
such a correlation. FR-I galaxies, on the other hand, have a much higher
source density and there exist several nearby sources which could be
responsible for a directional correlation. The most prominent FR-I
galaxies are Cen~A and M~87, as well as the flat spectrum radio
sources BL Lac and Perseus~A. Flat spectrum radio sources with low
luminosities, called BL Lac objects, are
typically interpreted as FR-I galaxies with their jet pointing
directly towards Earth. 

Observations of the jet structure of M~87~\citep{walker2008} and BL
Lac \citep{marscher2008} near the central black hole indicate that, while there can be
moving shocks between $10$~and~$1000$ Schwarzschild-radii, the first
strong, stationary shock occurs at $\sim 3000$ Schwarz\-schild-radii ($r_{g}$), as
already discussed by~\cite{markoff2001,markoff2005}. This is further
confirmed by \cite{britzen2008} in the case of the BL Lac type object S5~1803+784. At $\sim
3000\,r_{g}$, the optical depth
for proton interactions with photons from the disk is around 2\%. With
synchrotron photons from the jet as a target, the optical depth is $\sim 90\%$ at
the same distance from the central black hole. In both cases, the
optical depth decreases with the distance from the black hole,
$\tau_{p\,\gamma}\sim z_{j}^{-1}$.
 Therefore, we predict that neutrinos
are produced in the narrow jet, close to the central black hole. Their
emission is beamed due to the proton's highly
relativistic motion along the jet. Therefore, neutrinos can only be
observed for sources pointing directly towards Earth. Protons, on the
other hand, dominantly arise in the last strong shock of the jet, at
several kpc distance from the central black hole, where the optical
depth is close to zero.

Using the correlation between UHECRs and the distribution of AGN as a
measure for the Cosmic Ray flux from the supergalactic plane and
connecting this to the neutrino emission produced near the foot the
AGN jet results
in a predicted neutrino signal about an order of magnitude below the
current AMANDA limit~\cite{jess_diffuse}. IceCube will be able to test
this model within the first years of operation.

The general terms of this model are independent of Auger data, and could have been normalized to the established UHECR flux~\citep{gaisser_stanev2006}, assuming that 
radio galaxies are the sources~\citep[e.g.]{ginzburg_syrovatskii1963,biermann_strittmatter1987}.  Auger does confirm a correlation with 
the distribution of active galactic nuclei~\citep{auger_science2007,auger_astrop2008}, 
while HiRes~\citep{hires2008} does not, using the same energy 
threshold, and the same procedure.  It is clear that more data are 
required, and the energy threshold might ultimately be seriously 
different; the results show that a confirmation from a larger data set 
will be necessary.
\ack
%%%%%%%%%%%%%%%%%%%%%%%%%%%%%%%%%%%%%%%%%%%%%%%%%
%\subsection*{Acknowledgments}
JKB and PLB would like to thank Francis Halzen, Phil P.~Kronberg,
Dongsu Ryu, Todor Stanev, Paul Wiita, Ina Sarcevic, Wolfgang Rhode, John Belz and Markus Roth, as
well our IceCube and Auger collaborators for
inspiring discussions. PLB wishes to especially thank Ioana Du\c{t}an,
Laurentiu Caramete, Alexandru Curu\c{t}iu for work on the physics of
the sources of ultra high energy cosmic rays, which is in preparation now. Support for JKB is coming from the
DFG grant BE-3714/3-1 and from the IceCube grant BMBF (05 CI5PE1/0). Support for PLB is coming from the
AUGER membership and theory grant 05 CU 5PD 1/2 via DESY/BMBF, as well
as from VIHKOS. 
%%
%
%\bibliographystyle{elsart-harv}
%\bibliography{lib}

\end{document}